\documentclass[twocolumn,showpacs,aps,prl]{revtex4}
\usepackage{dcolumn}
\usepackage{epsf}
\begin{document}
\title{Origin of the efficient light emission from inversion
domain boundaries in GaN } 
\author{Vincenzo  Fiorentini}\email{vincenzo.fiorentini@dsf.unica.it}
\affiliation{INFM and Dipartimento di  Fisica, Universit\`a di Cagliari,
Cittadella  Universitaria, I-09042 Monserrato (CA), Italy}
\date{\today} 
\begin{abstract}
Intentionally-produced inversion domain boundaries in GaN
have been reported to be highly efficient shallow recombination
centers. Here I report a rationale for this 
phenomenon based 
on ab initio density-functional calculations. 
I also propose a model, based on the existence of 
polarization in GaN, of the observation that
a domain boundary acts as a rectifying junction
under voltage applied between the two opposite-polarity 
surfaces.
\end{abstract}
\pacs{61.72.Bb,  
      61.72.nn,  
      77.22.Ej}  

\maketitle
Nitride semiconductors, and GaN in particular,
are major competitors in the optoelectronics market 
at the high frequency end of the visible spectrum.
Understandably, considerable attention has been devoted to the
nature of   recombination centers in In-alloyed GaN,  both 
non-radiative (mainly threading dislocations) and radiative
(allegedly, In-rich regions). Not nearly as much work
exists on recombination centers in plain GaN.  
In recent times, 
extended defects originating from the formation of
polarity domain boundaries, and known as inversion domain boundaries
(henceforth IDB for brevity),
 have come into focus in this context. They have been
studied theoretically \cite{idb-th},  directly imaged by
 high-resolution TEM \cite{idb-tem} and piezoresponse force
microscopy \cite{idb-pfm}, and finally, 
 intentionally produced in MBE epitaxial growth on patterned
substrates \cite{idb-pl}. By means of integrated PL topography 
\cite{idb-pl} it was observed that, unexpectedly, these IDBs exhibit a
tenfold more intense light emission compared to plain Ga-face surface
 areas. The transition energy is about 30 meV below the normal emission,
and unlike the latter, it is polarized in the plane of the defect. The
 IDBs thus qualify as highly efficient  recombination
centers. The first purpose of this paper is to report a rationale
for  this phenomenon, based on ab initio density-functional
calculations. A further puzzling observation \cite{idb-rect} is that
an IDB between  metal contacts placed on the Ga-face and N-face
surface regions, acts as a rectifying junction (direct bias being Ga-face 
to N-face). Here, I also propose a
model of the latter behavior, based on the existence of permanent bulk
polarization in GaN \cite{pol}. 

Being a
 wurtzite crystal, GaN possesses a singular polar axis
 labeled (0001)   with the convention 
that the positive direction  points from the cation to the anion,
and an  attendant permanent polarization vector
oriented as  (000$\overline{\rm1}$). There exist  two  geometrically 
distinct crystal terminations, the (0001) or Ga-face   and
 (000$\overline{\rm1}$) or N-face, whereby a Ga (respectively, a N)
atom would have a single ``dangling bond'' orthogonal to the surface.
 Two regions of opposite polarity, conveniently   labeled
 Ga-face and N-face, meet at an IDB. 
There are at least two plausible models thereof, which were discussed 
earlier by Northrup {\it et al.} \cite{idb-th}.
 In the first, cations and anions are exchanged 
in half of the crystal, giving  rise to a geometrically undistorted 
interface, which however contains two (one N--N and one
 Ga--Ga)  wrong  
bonds per cell \cite{nota}. The second model, also known as 
IDB$^*$ and found in Ref.\onlinecite{idb-th} to be decidedly favored 
energetically over the first, is obtained  shifting one of the
 inverted half-cristals by half the cell length ($c$/2)
  along the singular axis. 
This enables all atoms to be correctly bonded, at the only cost of 
a modest distortion of the local bond angles, the largest single angle 
change being $\sim$109$^{\circ}$  to 90$^{\circ}$. 
In this work, I will be  considering the latter model. 

I calculate total energies, forces, stress, and electronic 
potential and density from  first-principles, within density functional 
theory \cite{dft}  in the Ceperley-Alder \cite{ca} local-density
approximation to exchange and correlation, 
using the accurate all-electron frozen-core  Projector-Augmented Waves (PAW) method \cite{paw} as  implemented
\cite{vasp-paw} in the Vienna Ab-initio Simulation Package (VASP)
\cite{vasp}. The plane wave cutoff is set at 290 eV. The partial wave
expansions for N and Ga include angular momenta up to $l$=2 and $l$=3,
respectively, with all on-site  projectors included. Ga 3$d$ electrons
are treated as valence.  Accurate reciprocal-lattice grids are used,
with no wrap-around aliasing allowed in Fourier transforms.  
Monkhorst-Pack k-space integration grids  are used; for the  
bulk, converged   results are obtained with the (4$\times$4$\times$4) 
mesh, tested  up to the (12$\times$12$\times$12). The resulting crystal 
structure of GaN is $a_{\rm GaN}$=3.160 \AA, $c$/$a_{\rm GaN}$ =1.626, 
$u$=0.3769, which matches closely (maximum deviation: 0.9 \% for
$a_{\rm GaN}$) the accepted experimental values $a_{\rm exp}$=3.189
\AA,  $c$/$a_{\rm exp}$=1.626, $u$=0.377 \cite{exp-gan} -- one of the
best descriptions of  bulk structure  ever achieved   for GaN. 

The IDB$^*$ planar defect 
can be simulated  in periodic boundary conditions via  repeated 
supercells. I use a crystalline GaN orthorombic supercell comprised 
of 24 formula units (48 atoms), with linear dimensions (1, 1.626, 6
$\sqrt{3}$) in units of  $a_{\rm GaN}$. After testing
the  ($n$$\times$1$\times$$n$) and  ($n$$\times$2$\times$$n$) 
k-point meshes with $n$=2,4, and 8, I used the
(4$\times$2$\times$4) for all the results reported here.
The  symmetry of the system  is
found to be almost exactly D$_{2h}$, although this symmetry is not
explicitly enforced. When it is, the k-space grid includes 4 special
points in the supercell Brillouin reduced zone; otherwise the mesh
comprises 8 points in the reduced zone. 
 
Before relaxation,  the formation energy of the IDB defect is 
40 meV/\AA$^2$. To allow for possible defect-induced 
strains, the cell length in the $y$  direction 
(orthogonal to the defect plane) was adjusted, with
the atomic geometries being constantly optimized  to obtain residual 
force components below 0.01 eV/\AA. The cell length 
does not change  within numerical accuracy; some local relaxation
occurs in the planes  immediately adjacent the defect, all  other
atomic positions being  unperturbed. Relaxation reduces the formation
energy to 
20 meV/\AA$^2$, a rather  small value compared to the 
114 meV/\AA$^2$ of two relaxed  (10$\overline{\rm 1}$0) surfaces of 
GaN \cite{nonpolar},  and agreeing well with the 25 meV/\AA$^2$ of 
Ref.\cite{idb-th}.

I calculate the transition energies, using the so called $\Delta$SCF
principle \cite{sham}, as the combination of (variational) total energies
of the system with varying number of electrons
\begin{equation}  
E_{\rm gap} = E_{\rm tot}(N+1) - 2 E_{\rm tot}(N) + E_{\rm
tot}(N-1).
\label{dscf}
\end{equation}
This expression is exact in the $N\rightarrow\infty$ limit,
but it turns out to be pretty accurate also at finite $N$
to estimate the gap of crystals \cite{nota2}. The added charge 
is compensated by a uniform background.
Since  part of the excess charge localizes 
in the defect plane, I apply a correction $\sigma^2\, 
d_{\rm cell}/2\varepsilon_r$ ($\sigma$ the interface charge,
$d_{\rm cell}$ the cell length, 
$\varepsilon_r$ the static dielectric permittivity)
to account for the spurious interaction of such sheet charges,
screened by the static dielectric response of GaN. This
 correction  is in the order of 10 $\mu$eV, i.e.
fully negligible on the scale of other
 computational approximations (e.g. frozen core, LDA, etc.) 
and convergence errors.
\begin{figure}[h]
\epsfclipon \epsfxsize=8cm \epsffile{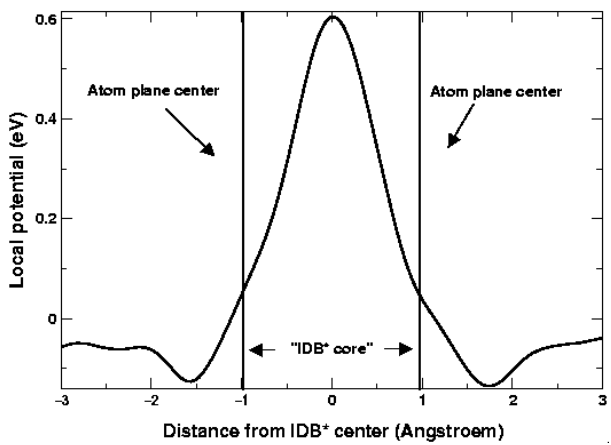}
\caption {Local self-consistent electronic potential at the IDB$^*$}
\label{mhat}
\end{figure}

The transition energy obtained for the bulk undefected cell is
3.49 eV, in good agreement with experiment. For the defected cell,
the transition energy is also 3.49 eV, thus predicting a red shift of
zero, compared to the experimental value of 30 meV. 
In the calculations the excess electron or hole are added separately
to the system, and they do not interact in any way, so that
  excitonic effect are not described.

However, in view of the low experimental temperature, it is 
fairly natural to suppose that the observed  shift may 
 be of excitonic origin.
Although electron-hole interaction is not included, it 
is possible to pinpoint the origin of the effect looking at the
precursors of the exciton, i.e. the electron and hole densities
and attendant self-consistent potential. Specifically, let us
examine in Fig. \ref{mhat} the self-consistent electronic potential 
of the defected cell in the neutral charge 
state (no extra charge), filtered through a macroscopic average
\cite{macro}  eliminating the periodic microscopic oscillations.
 As can be seen, the potential in the vicinity of the defect
is such as to  attract electrons with its weak but far-reaching 
side wings on the outer side of the defect core, and repelling them 
off the core itself; contrarywise, holes are attracted to the
defect core. Both electrons and holes are thus expected to localize in
the vicinity of the defect (which should enhance excitonic interactions).

\begin{figure}[h]
\epsfclipon \epsfxsize=8cm \epsffile{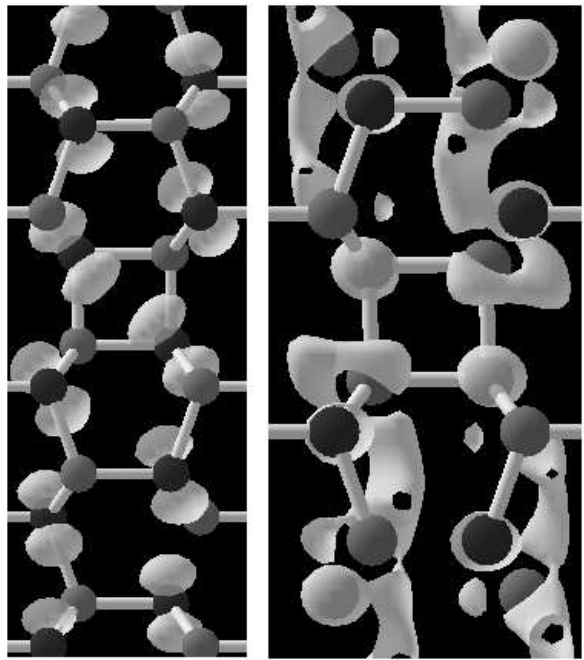}
\caption{Density of excess hole (left) and electron near the
IDB$^*$ defect.}
\label{e-h-dens}
\end{figure}

This is confirmed by inspection of the excess electron and hole charge
densities, obtained as density differences between charged and neutral
cells. As can be seen \cite{vaspview}
 in Fig. \ref{e-h-dens}, electrons accumulate
just outside the defect and onto the central repulsive wall, whereas
(although this is not especially clear from the picture)
the hole density is enhanced in the defect core. 
 
Inspection  also helps rationalize the observed emission
polarization in the defect plane. The electron density, mostly of
cation $s$ character, is ``planarized'' within the two-dimensional
lateral wells  of the defect potential. The hole density away from the
defect consists of a lobe along the $y$   direction: this
is because the valence top state in GaN is a $\Gamma_9$ doublet of
$p_xp_y$ nature (the $x$ lobe is not visible, as the $x$ direction
is orthogonal to the plane of the picture); near the defect the $p_y$ lobe 
pivots around the $x$ axis  admixing some character of the $p_z$ 
singlet $\Gamma_6$ which lies some 50 meV lower than $\Gamma_9$
 in unperturbed GaN.  Overall this results in an enhanced directional
overlap of the electron and hole functions.

\begin{figure}[h]
\epsfclipon \epsfxsize=8cm \epsffile{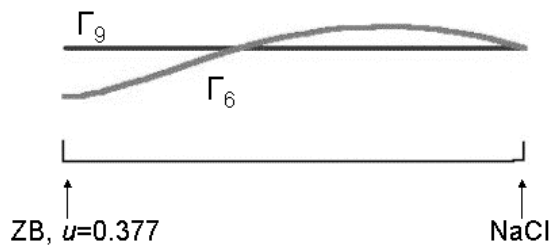}
\caption {Schematic of valence states admixture 
between wurtzite-like and NaCl-like environments.}
\label{zbtonacl}
\end{figure}

 A rationale for this admixture
can be extracted from the following model (refer to
Fig. \ref{zbtonacl}).
 Consider zincblende
GaN, with its three-fold degenerate  valence band top.
 Stretch a bond along one of the (111)-equivalent direction;
the degeneracy is removed, and as  in wurtzite the  
$\Gamma_6$ singlet is lower than the $\Gamma_9$ doublet. Now
displace, say, the anion along the chosen (111) direction, and monitor
the relative position of the states. What is observed 
 is the situation schematized in Fig. \ref{zbtonacl}: as the
distortion  progresses the $p_z$ singlet nears the doublet, becomes
higher in energy (i.e. the band top), then merges back with the
doublet into the threefold-degenerate   valence top of the NaCl
structure. The  IDB$^*$ embodies indeed a local NaCl-like geometry,
with one of the $a$-plane bonds at right angles with the polar-axis
bond. This  distortion is the cause of  the admixture of the two top
valence states. 


\begin{figure}[h]
\epsfclipon \epsfxsize=8cm \epsffile{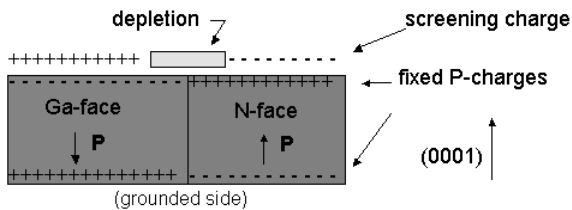}
\caption {Schematic of the polarization-driven 
rectifying mechanism.}
\label{polrect}
\end{figure}
Another  remarkable effect, observed by Stutzmann {\it 
et al.} \cite{idb-rect}, is that  an IDB$^*$  
acts as a rectifying junction when
electrically biased by two contacts placed respectively
 on the adjacent  Ga- and N-face regions, direct-bias being
 realized by the positive pole on the Ga-face region. This behavior
can be qualitatively explained by the existence of
an intrinsic macroscopic polarization in wurtzite \cite{pol}, as
schematized in Fig. \ref{polrect}. The two adjacent opposite-polarity
regions possess opposite intrinsic polarization vectors. This results
in fixed polarization charges at the upper and lower ends of the
sample, and changing sign across the IDB junction.  On the upper side,
mobile charges or impurities from  the vacuum neutralize the
polarization charges on each of  the regions. At the border between
the latter, the  compensating charges will mutually neutralize and
deplete, creating a junction-like potential which opposes the current
flow from Ga-face to N-face. Direct polarization with the positive
pole on the Ga-face side will then cause  a current flow from Ga-face
to N-face. It is difficult to say whether this current will  flow
subsurface or otherwise. Certainly, however,  the lower side of the
polarized layer,  a few $\mu$m below the surface, will not be involved. 
Thereby,   charges from, or defects in  the nucleation layer are
assumed to neutralize permanently the polarization charge.  

In summary I have given an interpretation of the observed 
enhanced light emission from inversion domain boundaries in GaN, and I have
suggested a qualitative explanation of the rectification effect
observed across the boundaries themselves.

I dedicate this work to the memory of my father Ugo Fiorentini.
Support from the Italian Ministry of Research under the PRIN 
funding programme (years 2000 and 2001) is acknowledged.

%

\begin{thebibliography}{99}
\bibitem{idb-th}
J. E. Northrup, J. Neugebauer, and L. T. Romano,
Phys. Rev. Lett. {\bf 77}, 103 (1996).
\bibitem{idb-tem}
C. Iwamoto, X. Q. Shen, H. Okumura, H. Matuhata, and
Y. Ikuhara, Appl. Phys. Lett. 79,  3941 (2001).
\bibitem{idb-pfm}
B. J. Rodriguez, A. Gruverman, A. I. Kingon, R. J.
Nemanich, and O. Ambacher,
 Appl. Phys. Lett. 80,  4166 (2002). 
\bibitem{idb-pl}
P. J. Schuck, M. D. Mason, R. D. Grober, O. Ambacher, A. P. Lima,
C. Miskys, R. Dimitrov, and M. Stutzmann, Appl. Phys. Lett. 79,  952
(2001). 
\bibitem{idb-rect}
M. Stutzmann, O.
Ambacher, M. Eickhoff, U. Karrer, A. P. Lima, R. Neuberger,
J. Schalwig, R. Dimitrov, P. J. Schuck, and R. D. Grober,
phys. stat. sol.(b) {\bf 228}, 505 (2001).
\bibitem{pol}
See F. Bernardini and V. Fiorentini,
Phys. Rev. B {\bf 64}, 085207 (2001), and
 A. Zoroddu, F. Bernardini, P. Ruggerone, and V. Fiorentini,
Phys. Rev. B {\bf 64}, 045208 (2001) for entries to the literature.
\bibitem{nota}
There is of course nothing wrong about Ga-Ga and N-N bonds in themselves;
here however the coordination forced by the surrounding structure
spoils any energy gain. 
\bibitem{dft}
R. M. Dreizler and E. K. U. Gross, {\it Density Functional
Theory} (Springer, Berlin 1988).
\bibitem{ca}
D. M. Ceperley and B. J. Alder,
Phys. Rev. Lett. {\bf 45}, 566-569 (1980)
\bibitem{paw}
P. E. Bl\"ochl, Phys. Rev. B {\bf 50}, 17953 (1994).
\bibitem{vasp-paw}
G. Kresse and D. Joubert, Phys. Rev. B {\bf 59}, 1758 (1999). 
\bibitem{vasp}
G. Kresse and J. Hafner, Phys. Rev. B {\bf 47}, R558 (1993); 
G. Kresse and J. Furthm\"uller, Comput. Mater. Sci. {\bf 6}, 15 (1996);
G. Kresse and J. Furthm\"uller, Phys. Rev. B {\bf 54}, 11169 (1996);
 http://cms.mpi.univie.ac.at/vasp/.
\bibitem{exp-gan}
M. Leszczynski, H. Teisseyre, T. Suski, I. Grzegory,
M. Bockowski, J. Jun, S. Porowski, K. Pakula, J. M. Baranowski,
C. T. Foxon, and T. S. Chen, Appl. Phys. Lett. {\bf 69}, 73  (1996).
\bibitem{nonpolar}
A. Filippetti, V. Fiorentini, G. Cappellini, and A. Bosin,
Phys. Rev. B {\bf 59}, 8026 (1999).
\bibitem{gap}
L. J. Sham and M. Schl\"uter, 
Phys. Rev. Lett. {\bf 51}, 1888 (1983).
\bibitem{macro}
A. Baldereschi, S. Baroni, and R. Resta,
Phys. Rev. Lett. {\bf 61}, 634 (1988);
M. Peressi, N. Binggeli, and A. Baldereschi, 
J. Phys. D {\bf 31}, 1273 (1998).
\bibitem{vaspview}
VASP Data Viewer, {\tt http://vaspview.sourceforge.net}.
\bibitem{sham}
J. P. Perdew and M. Levy, 
Phys. Rev. Lett. {\bf 51}, 1884 (1983);
L. J. Sham and M. Schl\"uter, {\it ibid.}, 1888;
Phys. Rev. B {\bf 32}, 3883 (1993).
\bibitem{nota2}
The rationale of this estimate \cite{sham} is that the 
addition (e.g.) of an electron to an $N$-electron system 
 corrects explicitly for the DFT exchange-correlation discontinuity across
the gap. This is an {\it o}(1) effect. The errors due to the 
finiteness of the systems are   {\it o}(1/N) and, while sizable
($\sim$0.2 eV) in the individual  charged-cell   calculations,
 cancel out almost completely from the final estimate.
Results of similar quality were obtained 
for AlN [A. Fara, F. Bernardini, and V. Fiorentini,
J. Appl. Phys. {\bf 85}, 2001 (1999)],  Ge [P. Delugas and
V. Fiorentini, unpublished], and Si [G. Lopez and
V. Fiorentini, unpublished]. 
\end{thebibliography}
\end{document}